\documentclass[aps,prb,twocolumn,groupedaddress,amsmath,amsfonts,showpacs]{revtex4}

\usepackage[dvipdfm]{graphicx}
\usepackage{graphicx}
\usepackage{bm} 
\usepackage{braket}
\usepackage[usenames]{color}
\usepackage{comment}

\begin{document}


\title{Dynamical axion in topological superconductors and superfluids}


\author{Ken Shiozaki}
\author{Satoshi Fujimoto}
\affiliation{Department of Physics, Kyoto University, Kyoto 606-8502, Japan}




\date{\today}

\begin{abstract}
We consider dynamical axion phenomena in topological superconductors and superfluids in three spatial dimensions in terms of the gravitoelectromagnetic topological action, 
in which the axion field couples with mechanical rotation under finite temperature gradient. 
The dynamical axion is induced by relative phase fluctuations between topological and $s$-wave superconducting orders. 
We show that an antisymmetric spin-orbit interaction which induces parity-mixing of Cooper pairs 
enlarges the parameter region in which the dynamical axion fluctuation appears as a low-energy excitation. 
We propose that the dynamical axion increases the moment of inertia, and in the case of ac mechanical rotation, i.e. 
a shaking motion with a finite frequency $\omega$,
as $\omega$ approaches the dynamical axion fluctuation mass, the observation of this effect becomes feasible. 
\end{abstract}



\maketitle

\section{Introduction}
\label{Intro}
Topological insulators and superconductors are characterized by momentum space topology.\cite{hasan2010colloquium, qi2011topological}
The band insulators and (fully-gapped) superconductors have no low-energy fermionic quasi-particle excitations in the bulk.
However, 
nontrivial global topology of the ground state in the momentum space yields robust gapless states at boundaries or defects, the existence of which is an important feature of topological phases. 
Furthermore, momentum space topology is also associated with electromagnetic or thermal polarization of these systems. 
For example, charge polarization \cite{king1993theory,goldstone1981fractional} is given by a momentum space integral of the Berry connection and couples with electric fields. 
Magnetoelectric polarization \cite{qi2008topological} (or called as the axion field) $\theta$ characterizing the momentum space topology of band insulators gives rise to 
axion electrodynamics \cite{wilczek1987two}, in which magnetoelectric polarization $\theta$ couples with the electromagnetic field via $\mathcal{L}_{\mathrm{top}} = \frac{\alpha}{4 \pi^2} \theta \bm{E} \cdot \bm{B}$.
Because $\bm{E} \cdot \bm{B}$ is a total derivative, physical phenomena associated with $\mathcal{L}_{\mathrm{top}}$ arise only when $\theta$ is spatially or temporally inhomogeneous. 
A domain wall of $\theta$ leads to the anomalous Hall effect and the image monopole effect. \cite{qi2009inducing}
Thermal or quantum fluctuations of the axion field $\theta$ also yield various interesting phenomena. 
For example, in a time-reversal invariant $\mathbb{Z}_2$ topological insulator in three dimension, inter-orbital antiferromagnetic fluctuations induce the dynamical axion field, 
which couples with electromagnetic fields, 
leading to an axionic polariton under an applied magnetic field \cite{li2010dynamical} and magnetic instability under an applied electrostatic field. \cite{ooguri2012instability}

In the case of topological superconductors (TSCs), since charge, and also spin if the spin-orbit interaction (SOI) presents, are not conserved, it is difficult to detect topological characters in electromagnetic responses. 
However, thermal responses can be good probes for topological nontriviality because energy is still conserved. 
TSCs with time-reversal symmetry (class DIII) in three-spatial dimensions are classified by an integer $\mathbb{Z}$, i.e. the so-called winding number, \cite{schnyder2008classification}
and the TSC characterized by the topological invariant $N$ possesses $N$ gapless Majorana fermions at open boundaries. 
If mass gaps are induced in surface gapless Majorana fermions, the (2+1)-dimensional surface exhibits the thermal Hall effect, 
which can be interpreted in the context of the thermal axion physics where the axion field $\theta$ equals to $N \pi$. \cite{ryu2012electromagnetic, wang2011topological, nomura2012cross}  From analogy to the dynamical axion in the topological insulators, \cite{li2010dynamical} 
one can deduce that for TSCs, the dynamical axion field in TSCs can be provided by imaginary $s$-wave superconducting fluctuations,
which break quantization of magnetoelectric polarization and lead to dynamics of $\theta$. \cite{shiozaki2013electromagnetic}

In the present paper, we discuss the axion physics in TSCs mainly focusing on dynamical effects, i.e. fluctuations of the axion field in TSCs. 
We investigate low-energy excitations of the dynamical axion using a concrete model of a $p$-wave TSC with an $s$-wave channel attractive interaction
which is necessary for inducing dynamics of axion. 
We also discuss the effect of the SOI in the case with noncentrosymmetric crystal structures, and show that the inversion-symmetry-breaking SOI enhances significantly dynamical axion fluctuations. We also discuss some implications for experimentally observable phenomena involving dynamical axion.

The organization of this paper is as follows. 
In Sec.\ref{TopTerm}, we first give topological $\theta$ terms that may be related to the axion physics in thermal (or gravitational) responses. 
In Sec. III, we would like to comment on the $\mathbb{Z}$ classification of TSCs and illustrate a difference from $\mathbb{Z}_2$ classification. 
The relation between the imaginary $s$-wave superconducting fluctuations and dynamical axion is clarified. 
In Sec.\ref{Dy-Axion}, we consider a concrete microscopic model of a TSC in which the dynamical action appears  
at a sufficiently low-energy scale compared to the bulk superconducting gap, and clarify the effect of the SOI raised by broken inversion symmetry, which makes low-energy dynamical axion more feasible. 
In Sec.\ref{Dy-Phe}, we discuss experimentally observable phenomena driven by the dynamical axion in TSC. 
We conclude in Sec.\ref{Conc} with some remarks.

\section{Gravitational topological action term for topological superconductors}
\label{TopTerm}
In this section, we briefly review previous argument on topological action terms and effective internal energy related with thermal responses in TSCs, which are
the basis for dynamical axion phenomena. 
In TSCs, a conserved quantity is energy, and hence thermal responses can be used for probing the topological nontriviality. 
For translational invariant superfluids, momentum is also a conserved current. 
Probe fields which couple with energy-momentum tensor are metric. 
Here we consider topological action terms which are constructed from the metric degrees of freedom, 
These action terms are known as the gravitational instanton term in (3+1) dimensions and the gravitational Chern-Simons (CS) term in (2+1) dimensions. 
Also, we introduce another type of a topological term which is related with the gravitoelectromagnetism and the torsional anomaly. \cite{hidaka2013viscoelastic, hughes2013torsional}

\subsection{(2+1) dimensions}
\label{}
In two-dimensional systems, TSCs characterized by the nontrivial Chern number have the gravitational CS 3 form as a low-energy effective theory, \cite{volovik2003universe, read2000paired}
\begin{equation}\begin{split}
S_{CS} = \frac{1}{4 \pi} \frac{c}{24} \int d^3 x \epsilon^{\mu\nu\rho} \mathrm{tr} \left( \omega_{\mu} \partial_{\nu} \omega_{\rho} + \frac{2}{3} \omega_{\mu} \omega_{\nu} \omega_{\rho} \right) 
\end{split}
\label{gCS}
\end{equation}
with $x=(c_0 t, \bm{r})$ where $c_0$ is the Fermi velocity, $\omega_{\mu}$ the spin connections determined by the metric and $c = 1/2$, which is the central charge of
the Ising conformal field theory. It has been discussed that there should be an edge channel with the central charge $c=1/2$, which raises the thermal Hall effect in Hall-bar geometry. \cite{read2000paired}
It is important to note that the gravitational CS term (\ref{gCS}) does not directly lead to the thermal Hall effect in (2+1) dimension, 
because (\ref{gCS}) yields the energy current proportional to gradient of the Ricci tensor which corresponds to a spatially second order differential of a temperature field. \cite{stone2012gravitational}
In this respect, the gravitational CS action differs from the electromagnetic CS action $S_{CS} = c \frac{\alpha}{4 \pi} \int d^3 x \epsilon^{\mu\nu\rho} A_{\mu} \partial_{\nu} A_{\rho} $, 
which yields the Hall current proportional to electric field via a functional derivative, $j_y = c (e^2/h) E_x$. 

On the other hand, a careful calculation of the thermal Hall effect including a contribution of the dia-thermal current 
(a change of the local thermal current operator due to an applied gravitational field), \cite{smrcka1977transport, nomura2012cross}
or equivalently, a contribution from the thermal magnetization current, \cite{qin2011energy, sumiyoshi2013quantum}
provides the generalized Wiedemann-Franz low at low temperature, which suggests the thermal Hall effect in the (2+1) dimensional bulk system with a topological thermal coefficient defined by $j_{Hy} = \kappa_H \partial_x T$, 
\begin{equation}\begin{split}
\kappa_H = c \frac{\pi^2 k_B^2 T}{3 h}. 
\end{split}\label{KH}\end{equation}
This thermal Hall conductivity agrees with the result  obtained from the (1+1) dimensional edge theory. 
Here, we would like to note that an important role of the energy magnetization contribution can be seen in various systems. 
For instance, let us consider massive Weyl fermions described by the bulk Hamiltonian $\mathcal{H} =  \sigma_1 k_x + \sigma_2 k_y + m \sigma_3$.
In this system, the standard Kubo formula for energy current correlation functions gives no contribution, and 
the thermal Hall conductivity $\kappa_{H}$ stems only from the energy magnetization. 

The topological thermal Hall conductivity (\ref{KH}) from the (2+1) dimensional bulk calculation suggests an alternative low energy effective action describing the thermal Hall effect, 
\begin{equation}\begin{split}
S_{CS} = c \frac{\pi^2 k_B^2 T^2}{6 h} \int dt d^2 \bm{r} \epsilon^{\mu\nu\rho} A_{E,\mu} \partial_{\nu} A_{E,\rho}
\end{split}\label{gEMCS}\end{equation}
where $A_{E\mu} = (A_{0E}, \bm{A}_E)$ couple with an energy density $h(\bm{x})$ and an energy current density $\bm{j}_E(\bm{x})$ via $H_c = \int d^2 x \left( h(\bm{x}) A_{E0} - \bm{j}_E(\bm{x}) \cdot \bm{A}_E \right)$. 
In the action (\ref{gEMCS}), an overall coefficient has a dimension of the square of energy, which is the same as the the torsional anomaly. \cite{hughes2013torsional}
The relation between the thermal Hall effect and the torsional anomaly was discussed by Hidaka et al. 
at zero temperature.\cite{hidaka2013viscoelastic} 
Also, 
Shi and Cheng
discussed\cite{shi2012heat} the gauge invariance for $\bm{A}_E$ within a first order of $A_{E\mu}$ 
by using the scaling relation for the energy current density under gravitational potential. \cite{cooper1997thermoelectric, qin2011energy}
We could not derive the action (\ref{gEMCS}) directly from microscopic Hamiltonian of electron systems coupled with gravitoelectromagnetic fields. 
However, we deduce it from the analogy with the relation between the (2+1) dimensional electromagnetic CS action and the electric Hall conductivity.
It is still an important open issue to establish 
the low-energy effective field theory of the thermal Hall effect. 


\subsection{(3+1) dimensions}
For (3+1) dimensional TSCs with time-reversal symmetry classified by the topological invariant $N$, a low-energy gravitational effective theory 
is described by the gravitational instanton term, 
\begin{equation}\begin{split}
S_{\theta} = \frac{\theta}{1536 \pi^2} \int d^4 x \epsilon^{\mu\nu\rho\sigma} R^{\alpha}_{\beta\mu\nu} R^{\beta}_{\alpha\rho\sigma}
\end{split}\label{gTop}\end{equation}
with $\theta=\pi$ (mod $2\pi$).
Thus, Eq. (\ref{gTop}) provides a $\mathbb{Z}_2$ characterization of the TSCs classified by the parity of  the integer topological invariant $N$. 
For heterostructure geometry composed of the TSC and a trivial insulator (here, ``trivial" means the zero axion angle), 
when interface Majorana fermions are completely gapped out, the action (\ref{gTop}) leads to the interface half integer gravitational CS action 
\begin{equation}\begin{split}
S_{CS} 
&= \frac{1}{2} \frac{1}{4 \pi} \frac{c}{24} \int d^3 x \epsilon^{\mu\nu\rho} \mathrm{tr} \left( \omega_{\mu} \partial_{\nu} \omega_{\rho} + \frac{2}{3} \omega_{\mu} \omega_{\nu} \omega_{\rho} \right)
\end{split}\label{gCS2}\end{equation}
with $c = (N+2M)/2$ where $M$ is an integer depending on the microscopic structure of the interface, and determined by the signs of mass gaps of $N$ Majorana fermions. \cite{wang2011topological, shiozaki2013electromagnetic}
It is noted that as in the case of the pure (2+1) gravitational CS action (\ref{gCS}), the action (\ref{gCS2}) does not directly lead to the bulk thermal Hall effect. 

A more preferable action describing the interface thermal Hall effect is the following gravitoelectromagnetic $\theta$ term, \cite{nomura2012cross}
\begin{equation}\begin{split}
S_{\theta} = \theta \frac{\pi k_B^2 T^2 }{12 h} \int d t d^3 \bm{r} \bm{E}_E \cdot \bm{B}_E, 
\end{split}\label{gEMTop}\end{equation}
where $\bm{E}_E$ and $\bm{B}_E$ couple with the energy polarization and the energy magnetization, respectively. 
The topological action (\ref{gEMTop}) describes a thermal topological magnetoelectric effect, i.e., 
the energy magnetization induced by the gravitational field: $\bm{M}_E = \theta \frac{\pi k_B^2 T^2 }{12 h} \bm{E}_E $, 
and the energy polarization induced by the gravitomagnetic field; $\bm{P}_E = \theta \frac{\pi k_B^2 T^2 }{12 h} \bm{B}_E $. \cite{nomura2012cross} 
If there is a U(1) gauge structure in $\bm{E}_E, \bm{B}_E$, as in the case of the axion electrodynamics, \cite{qi2008topological}
a low-energy effective action for Majorana fermions at the interface between the TSC and the trivial insulator is given by, 
\begin{equation}\begin{split}
S_{CS} =  \frac{c}{2} \frac{\pi^2 k_B^2 T^2}{6 h} \int d^3 x \epsilon^{\mu\nu\rho} A_{E,\mu} \partial_{\nu} A_{E,\rho}, 
\end{split}\label{gEMCS2}\end{equation}
where $c = (N+2M)/2$. 


Here we sketch the derivation of the gravitoelectromagnetic topological internal energy presented by Nomura et al. \cite{nomura2012cross}
The time reversal symmetric (class DIII) TSC classified by topological invariant $N$ possesses $N$ gapless Majorana modes localized at the boundary
of the system. 
If mass gaps are induced in Majorana modes, each Majorana fermion gives rise to the half-integer thermal Hall effect described by 
the Hall conductivity $\kappa_H = \mathrm{sgn}(m) \frac{\pi^2 k_B^2 T}{12 h}$ where $m$ is the mass. 
They proved that the topological part of the internal energy for a TSC in a cylindrical geometry with a surface perturbation inducing mass gaps of the Majorana fermions is expressed as, 
\begin{equation}\begin{split}
U_{\mathrm{top}}
&= - \int d^3 \bm{r} \frac{k^2_{\mathrm{B}} T}{12 \hbar v^2} \theta \bm{\nabla} T \cdot \bm{\Omega}. 
\end{split}\label{TopFE}\end{equation}
Here $v$ is a velocity of surface Majorana fermion and $\theta$ is a constant determined by the sign of mass gaps of the surface Majorana fermions.  
They obtained this result from the calculations of the cross-correlated type response on the surface quasi-Lorentz symmetric system: 
(i) an angular momentum induced by the temperature gradient along $z$-axis, and 
(ii) heat population induced by mechanical rotation.

Eqs. (\ref{gTop}), (\ref{gEMTop}) and (\ref{TopFE}) imply that dynamics of the axion field $\theta$ give rise to dynamical thermal (gravitational)
responses. We investigate such dynamical axion effects of TSCs in the following.

\section{Basic features of class DIII topological superconductors}

\subsection{$\mathbb{Z}$ characterization of class DIII topological superconductors and gravito magnetoelectric polarization}
In this section, we summarize some basic properties of class DIII TSCs relevant to the following argument.
The topological classification of class DIII TSCs is given by integers $\mathbb{Z}$.  However, 
the gravitational instanton term (\ref{gTop})  describes only $\mathbb{Z}_2$ part, i.e., the parity of topological invariant.
Also, the internal energy (\ref{TopFE}) is derived by assuming a priori the existence of $N$ Majorana fermions on the surface.
As in the case of the magnetoelectric polarization in topological insulators \cite{qi2008topological}, 
in real systems, a non-topological even integer part of $\theta$ depends on the energy gap structure of the surface Majorana fermions. 
Hence it is possible to obtain the thermal conductivity exactly proportional to topological invariant $N$ by using a specific perturbation which generates
mass gap $m_i$ of the $i$-th Majorana fermion satisfying $\sum_{i} \mathrm{sgn}(m_i)= N$. 
For this perturbation, the (gravito) magnetoelectric polarization $\theta = N \pi$. 
Actually, class DIII TSCs inherently have the appropriate perturbation that can be constructed from a chiral symmetric structure of class DIII TSCs: $\Gamma \mathcal{H}_{\mathrm{BdG}} \Gamma^{-1} = - \mathcal{H}_{\mathrm{BdG}}$ 
where $\Gamma$ is determined by the combination of a time-reversal transformation and a particle-hole transformation. 
This perturbation is expressed as
\begin{equation}\begin{split}
V_{\Gamma} = \frac{\gamma}{2} \int d^3 x \Psi^{\dag}(\bm{x}) \Gamma \Psi(\bm{x})
\end{split}\label{GammaP}\end{equation}
where $\Psi(\bm{x})$ is the Nambu spinor.
The perturbation (\ref{GammaP}) induces the thermal Hall conductivity characterized by the topological invariant $N$ as $\kappa_H = N \frac{\pi^2 k_B^2 T}{12 h}$. \cite{wang2011topological, shiozaki2013electromagnetic}
To see the physical meaning of (\ref{GammaP}), let us consider a time-reversal symmetric superconductor. 
If we choose the Nambu representation $\Psi = (\psi_{\uparrow }, \psi_{\downarrow }, \psi^{\dag}_{\downarrow }, -\psi^{\dag}_{\uparrow })^T$, 
the time-reversal transformation $\Psi \mapsto \Theta \Psi$ is given by $\Theta = i \sigma_2 K$
and the particle-hole transformation $\Psi \mapsto \Xi \Psi$ is given by $\Xi = \tau_2 \sigma_2 K$ where $K$ is a complex conjugate operator. 
In this Nambu bases, $\Gamma = \tau_2$ (overall sign is arbitrary), then $V = - i \gamma \int d^3 x (\psi_{\uparrow }\psi_{\downarrow } - \psi^{\dag}_{\downarrow }\psi^{\dag}_{\uparrow }) $.
Hence $V$ is the imaginary 
$s$-wave pairing order. \cite{wang2011topological, shiozaki2013electromagnetic} 
Here, we have assumed that the global phase of the bulk pairing gap is fixed to be zero. From an effective description of the surface Majorana fermions under the perturbation of $V_{\Gamma}$, we can get $\kappa_H = N \frac{\pi^2 k_B^2 T}{12 h}$. 

Also, $\sum_{i} \mathrm{sgn}(m_i)= N$ can be derived from the surface jump of (gravito) magnetoelectric polarization defined by the Chern-Simons 3 form, 
\begin{equation}\begin{split}
\theta_{\mathrm{CS}} 
&= 2 \pi \int CS_3(\mathcal{A}) 
= \frac{1}{4 \pi} \int \mathrm{tr} \left( \mathcal{A} d \mathcal{A} + \frac{2}{3} \mathcal{A}^3 \right), 
\end{split}\label{axion}\end{equation}
where $\mathcal{A} = \mathcal{A}_{nm}(\bm{k}) = \Braket{u_n(\bm{k}) | d_{\bm{k}} u_m(\bm{k})}$ is the Berry connections 
determined by quasiparticle states of a bulk Bogoliubov-de Gennes (BdG) Hamiltonian $\mathcal{H}_{\mathrm{BdG}}(\bm{k}) \Ket{u_n(\bm{k})}= -E_n(\bm{k}) \Ket{u_n(\bm{k})}$. 
An adiabatic treatment of the surface structure between the TSC and the $V_{\Gamma}$ perturbation yields an adiabatic Hamiltonian 
$\mathcal{H}_{\mathrm{BdG}}(\bm{k}, \lambda) = (1-\lambda) \mathcal{H}_{\mathrm{BdG}}(\bm{k}) + \lambda \gamma \Gamma $ ($\lambda \in [0,1]$), 
and gives the surface jump of (gravito) magnetoelectric polarization by $\int_{\lambda=0}^{\lambda=1} d \theta_{\mathrm{CS}}(\lambda) = \mathrm{sgn} (\gamma) N \pi $. \cite{shiozaki2013electromagnetic}
Although the magnetoelectric polarization $\theta_{\mathrm{CS}}$ is gauge invariant only modulo $2 \pi$, 
the line integral of a small difference of $\theta_{\mathrm{CS}}(\lambda)$ is fully gauge invariant.
The derivation of  the topological internal energy (\ref{TopFE}) given by Nomura et al. is based on the surface half-integer thermal Hall effect, 
which limits the applicability of (\ref{TopFE}) to the case with a quantized value of $\theta$. 
However, the above observation that the $\mathbb{Z}$-part of $\theta$ coincides with the surface jump of the (gravito) magnetoelectric polarization $\theta_{\mathrm{CS}}$ defined by (\ref{axion}) strongly suggests that
non-quantized $\theta$ is given by $\theta_{\mathrm{CS}}$ as in the case of the axion electrodynamics. 
In this paper, we {\it assume} the topological action 
\begin{equation}\begin{split}
S_{\mathrm{top}}
&= - \int d t \int d^3 x \frac{k^2_{\mathrm{B}} T}{12 \hbar v^2} \theta_{\mathrm{CS}} \bm{\nabla} T \cdot \bm{\Omega} \\
\end{split}\label{FE}\end{equation}
for {\it non-quantized} value of $\theta_{\mathrm{CS}}$, and consider phenomena raised by the dynamical axion in TSCs. 
Later, we identify $\theta$ with $\theta_{\mathrm{CS}}$. 

Here, it is worth while mentioning the physical meaning of the non-quantized value of $\theta$. 
If the surface jump of $\theta$ is $\Delta \theta$, then the surface thermal Hall conductivity $\kappa_{H}$ is given by $\kappa_{H} = \Delta \theta \frac{\pi k_{\mathrm{B}}^2 T}{12 h}$. 
In the case that $\Delta \theta/\pi$ is an integer, 
the surface thermal Hall conductivity can be understood as contributions of surface massive Majorana fermions which gives $\kappa_H = \frac{\mathrm{sgn}(m)}{2} \frac{\pi^2 k_{\mathrm{B}}^2 T}{6 h}$, respectively. 
However, non-quantized values of $\kappa_H$ can not be derived from a pure (2+1)-dimensional surface effective theory. 
In fact, this comes from (3+1)-dimensional bulk wave functions.

\subsection{Difference between imaginary $s$-wave paring and Zeeman perturbations}

Before proceeding to the argument on dynamical axion, in this section, we clarify an important role of the chiral-symmetry breaking perturbation (\ref{GammaP}), i.e. imaginary $s$-wave pairing in TSCs.
Class DIII TSCs possess three types of symmetry, i.e. time-reversal symmetry $\Theta$, particle-hole symmetry $\Xi$, and chiral symmetry.
Since chiral symmetry is equivalent to the product of $\Theta$ and $\Xi$, one may expect that chiral-symmetry breaking perturbations
are essentially the same as time-reversal-symmetry breaking perturbations in superconductors where the particle-hole symmetry is preserved by definition. However, this naive expectation is not correct.
Indeed, we demonstrate here that  the chiral-symmetry breaking perturbation, i.e. the imaginary $s$-wave pairing, is indispensable for
realizing topological thermal responses characterized exactly by the winding number $N$, and that
time-reversal-symmetry breaking perturbations such as magnetic fields fail to give correct topological responses in certain  cases. 
For this purpose, we exploit a toy model for TSC BdG Hamiltonian with two-orbitals and spin $1/2$ internal degrees of freedom, 
\begin{widetext}
\begin{equation}\begin{split}
\mathcal{H}_{\mathrm{BdG}} 
&= 
\begin{pmatrix}
\left( \frac{\hbar^2 k^2}{2m} - \mu \right) \tau_3 + \frac{\Delta}{k_F} (k_x \sigma_1 + k_y \sigma_2 + k_z \sigma_3) \tau_1 & 0 \\
0 & \left( \frac{\hbar^2 k^2}{2m} - \mu \right) \tau_3 + \frac{\Delta}{k_F} (k_x \sigma_1 - k_y \sigma_2 - k_z \sigma_3) \tau_1 \\
\end{pmatrix}
\end{split}\end{equation}
\end{widetext}
where we take the Nambu spinor 
$\Psi = (\psi_{1\uparrow }, \psi_{1\downarrow }, \psi^{\dag}_{1\downarrow },-\psi^{\dag}_{1\uparrow }, \psi_{2\uparrow }, \psi_{2\downarrow }, \psi^{\dag}_{2\downarrow },-\psi^{\dag}_{2\uparrow })^T$. 
For simplicity, we assume that the fermion mass $m$ and the amplitude of the order parameter $\Delta$, and the Fermi wave number $k_F$ 
are the same for two orbitals, and 
there is no off-diagonal coupling between the orbital 1 and 2.
Since topological responses are not affected by details of microscopic systems, this assumption does not spoil the generality of the following result. From 
the global U(1) gauge arbitrariness of fermions, $\Delta$ can be fixed to be a positive value $\Delta>0$ without loss of generality. 
If $\mu>0$, $\mathcal{H}_{\mathrm{BdG}}$ describes a TSC with the topological invariant $N=2$ defined by 
\begin{equation}\begin{split}
N = -\frac{1}{48 \pi^2} \int \mathrm{tr} \ \Gamma \left[ \mathcal{H}_{\mathrm{BdG}}^{-1}(\bm{k}) d \mathcal{H}_{\mathrm{BdG}}(\bm{k}) \right]^3 , 
\end{split}\end{equation}
where $\Gamma = \tau_2$. 
On the other band, for $\mu<0$, the system is trivial with $N=0$.

Let us consider the surface Majorana modes of this system.
Suppose half-infinite geometry such that the TSC exists in the region $z<0$ and the region of $z>0$ is vacuum. We also assume that $\Delta$ is constant near the boundary at $z=0$. 
Under the fixed boundary condition for the wave function $\Phi(z=0) = 0$, 
the wave function of the gapless Majorana mode is given by 
\begin{equation}\begin{split}
\Phi_{1,i}(z) \sim
u_{1,i} e^{\tilde z} \mathrm{sinh} \left( \tilde z \sqrt{1-\tilde \mu(k_x,k_y)} \right) . 
\end{split}\end{equation}
Here $\tilde z = m \Delta z/\hbar^2 k_F$, $\tilde \mu(k_x,k_y) = 2 \hbar^2 k_F^2 \big( \mu-\hbar^2 (k_x^2+k_y^2)/2m \big) / m \Delta^2$, 
and $u_{1,i} (i=1,2)$ is the spinor for the helical Majorana fermions for the orbital 1, and we take 
$u_{1,1} = (1,0,i,0)^T $, $u_{1,2} = (0,i,0,1)^T $ in the basis of spin and particle-hole degrees of freedom, 
$\left( 
\Ket{
\begin{array}{ll}
\tau_3 = 1 \\
\sigma_3 = 1
\end{array} }, 
\Ket{
\begin{array}{ll}
\tau_3 = 1 \\
\sigma_3 = -1
\end{array} }, 
\Ket{
\begin{array}{ll}
\tau_3 = -1 \\
\sigma_3 = 1
\end{array} }, 
\Ket{
\begin{array}{ll}
\tau_3 = -1 \\
\sigma_3 = -1
\end{array} }
\right) $
in the space of the orbital 1. 
The Bogoliubov-de Gennes Hamiltonian $\hat{\mathcal{H}}_{\mathrm{BdG}}(k_x,k_y)$ is expressed in the Hilbert space spanned by the surface Majorana fermions $\Phi_1(z) = (\Phi_{1,1}(z), \Phi_{1,2}(z))$, 
\begin{equation}\begin{split}
\hat{\mathcal{H}}_{\mathrm{BdG}}(k_x,k_y) \Phi_1(z) = \Phi_1(z) \left[ \frac{\Delta}{k_F} (k_x \tilde \sigma_1 + k_y \tilde \sigma_2 ) \right], 
\end{split}\end{equation}
where, the Pauli matrices $\tilde{\bm{\sigma}} = (\tilde \sigma_1, \tilde \sigma_2, \tilde \sigma_3)$ are defined for the space spanned by $u_{1,1}$ and $u_{1,2}$. 
Similarly, for the orbital 2, the wave function of the gapless Majorana mode is given by 
\begin{equation}\begin{split}
\Phi_{2,i}(z) 
\sim u_{2,i} e^{\tilde z} \mathrm{sinh} \left( \tilde z \sqrt{1-\tilde \mu(k_x,k_y)} \right)
\end{split}\end{equation}
with $u_{2,1} = (i,0,1,0)^T$, $u_{2,2} = (0,1,0,i)^T$, and 
the Bogoliubov-de Gennes Hamiltonian $\hat{\mathcal{H}}_{\mathrm{BdG}}(k_x,k_y)$ is expressed in the Hilbert space spanned by the surface Majorana fermions $\Phi_2(z) = (\Phi_{2,1}(z), \Phi_{2,2}(z))$, 
\begin{equation}\begin{split}
\hat{\mathcal{H}}_{\mathrm{BdG}}(k_x,k_y) \Phi_2(z) = \Phi_2(z) \left[ -\frac{\Delta}{k_F} (k_x \tilde \sigma_1 - k_y \tilde \sigma_2 ) \right]. 
\end{split}\end{equation}
The surface gapless Majorana modes $\Phi_{1/2}(z)$ have opposite chiralities and different structures of the internal degrees of freedom, which come from a difference of the sign of $\Delta/k_F \tau_1 \sigma_3 (-i \partial_z)$. 
There are only two possible perturbations which can induce mass gap of gapless Majorana fermions : 
one is a Zeeman field perpendicular to the surface $h_z \sigma_3$, and the other one is an imaginary $s$-wave pairing order $\Delta^{\mathrm{Im}}_s \tau_2$. 
In the presence of both fields $h_z \sigma_3$ and $\Delta^{\mathrm{Im}}_s \tau_2$, the effective surface BdG Hamiltonians for the orbital 1 and 2 are, respectively, 
\begin{equation}\begin{split}
\mathcal{H}_{\mathrm{surf},1} = \frac{\Delta}{k_F} (k_x \tilde \sigma_1 + k_y \tilde \sigma_2 ) + ( h_z + \Delta^{\mathrm{Im}}_s) \tilde \sigma_3 
\end{split}\end{equation}
\begin{equation}\begin{split}
\mathcal{H}_{\mathrm{surf},2} = -\frac{\Delta}{k_F} (k_x \tilde \sigma_1 - k_y \tilde \sigma_2 ) + ( h_z - \Delta^{\mathrm{Im}}_s) \tilde \sigma_3. 
\end{split}\end{equation}
Then we get the thermal Hall conductivity, 
\begin{equation}\begin{split}
\kappa_{\mathrm{H}}(z=0)
&= \frac{\pi k_{\mathrm{B}}^2 T}{12 h} \left[ \mathrm{sgn}(h_z + \Delta^{\mathrm{Im}}_s) - \mathrm{sgn}(h_z - \Delta^{\mathrm{Im}}_s) \right]. 
\end{split}\end{equation}
Now, the role of $\Gamma$-perturbation (\ref{GammaP}) is obvious. 
In the case of the Zeeman field, there is no net thermal Hall currant. 
On the other hand, 
the imaginary $s$-wave pairing induces the same contributions from each surface Majorana fermions, 
which sum up to the thermal Hall conductivity with the coefficient equal to the topological invariant $N=2$.

The above discussions are also applicable to the cases where gapless Majorana fermions occur at time-reversal symmetric Dirac points of momentum space besides $\bm{k} = \bm{0}$. 
Thus, we have demonstrated that in the cases where spin structures of superconducting order parameters are different between different orbital bands, 
the imaginary $s$-wave perturbation and the Zeeman field perturbation give different results.
Only the former successfully gives the correct universal surface thermal Hall conductivity described by the bulk topological invariant $N$, 
while the latter fails to give topological thermal responses.

\subsection{Preliminary for dynamical axion : non-quantized axion angle for a $p$-wave TSC with an imaginary $s$-wave pairing order}
As mentioned previously, the imaginary $s$-wave pairing order in spin-$1/2$ systems, or generally, the $\Gamma$-perturbation ($\ref{GammaP}$) induces the change of the axion angle by a quantized value $\theta = N \pi$. 
Hence, thermal and quantum fluctuations of the imaginary $s$-wave superconducting order $\delta \Delta^{\mathrm{Im}}_s$ may give rise to dynamical axion $\delta \theta$ in TSCs. 
One may naively expect that the opposite may be also possible, i.e., the fluctuation of the imaginary topological superconducting order $\delta \Delta^{\mathrm{Im}}_{p}$ in the ordinary $s$-wave superconductor 
raises the dynamical axion. 
However, actually, dynamical axion in the latter case is much suppressed compared to the former case.
To see this, 
here, we consider a concrete example, the axion angle for the topological $p$-wave superconducting order
coexisting with the imaginary $s$-wave order described by 
\begin{equation}\begin{split}
\hat \Delta = \Delta_p \frac{\bm{k}}{k_F} \cdot \bm{\sigma} (i \sigma_2) + i \Delta_s^{\mathrm{Im}} (i \sigma_2), 
\end{split}\end{equation}
where $\Delta_p$ and $\Delta_s^{\mathrm{Im}}$ are real constants. 
The corresponding bulk BdG Hamiltonian is 
\begin{equation}\begin{split}
\mathcal{H}_{BdG}
&= \left( \frac{k^2}{2m}-\mu \right) \tau_3 + \tau_1 \Delta_p \frac{\bm{k}}{k_F} \cdot \bm{\sigma} - \tau_2 \Delta_s^{\mathrm{Im}} \\
&= \epsilon_F \left[ \left( \frac{k^2}{k_F^2} - 1 \right) \tau_3 + \tau_1 \frac{\Delta_p}{\epsilon_F} \frac{\bm{k}}{k_F} \cdot \bm{\sigma} - \tau_2 \frac{\Delta_s^{\mathrm{Im}}}{\epsilon_F}  \right], 
\end{split}\label{HBdG}\end{equation}
where we choose the Nambu spinor as $\Psi = (\psi_{\uparrow }, \psi_{\downarrow }, \psi^{\dag}_{\downarrow }, -\psi^{\dag}_{\uparrow })^T$. 
As shown in the second line of (\ref{HBdG}), the axion angle depends on the two parameters 
$\left( \Delta_p/\epsilon_F,\Delta_s^{\mathrm{Im}}/\epsilon_F \right)$ 
since the axion angle defined by (\ref{axion}) is independent of an overall factor of the BdG Hamiltonian and a scale of the wave number $\bm{k}$ for continuum models. 
It is noted that the latter simplification is not applicable to lattice systems. 
Parameterizing $\left( \Delta, \phi \right)$ by $\left( \Delta_p ,\Delta_s^{\mathrm{Im}} \right) = \left( \Delta \cos \phi, \Delta \sin \phi \right)$, 
we calculate the change of $\theta$ as a function of $\phi$ with a fixed value of $\Delta$, 
\begin{equation}\begin{split}
\frac{d \theta}{d \phi}
&= \frac{1}{16 \pi} \int d^3 k \ \epsilon^{\mu\nu\rho\sigma} \mathrm{tr} \left[ \mathcal{F}_{\mu\nu}(\bm{k},\phi) \mathcal{F}_{\rho\sigma}(\bm{k},\phi) \right] \\
&= \int_0^{\infty} d k \frac{3 \alpha ^4 k^2 \cos ^2\phi \left\{3 k^2+1-\left(k^2-1\right) \cos (2 \phi )\right\}}{2 \left(\alpha ^2 \sin ^2\phi+\alpha ^2 k^2 \cos^2 \phi+\left(k^2-1\right)^2\right)^{5/2}} 
\end{split}\end{equation}
with $\alpha = \Delta/\epsilon_F$. 
\begin{figure}[h]
\centering
  \includegraphics[width=\linewidth]{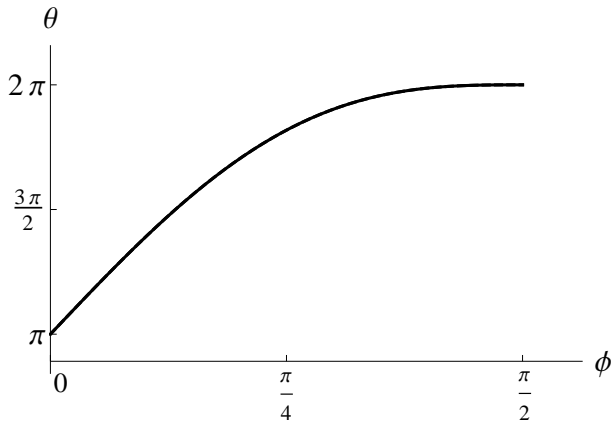}
 \caption{The axion angle $\theta$ as a function of $\phi$ for fixed gap amplitudes $\Delta/\epsilon_F= 0.003, \ 0.01, \ 0.1$. 
The three lines almost overlap each other. }
 \label{FIG1}
\end{figure}
In Fig.\ref{FIG1}, we show the axion angle $\theta$ as a function of $\phi$ for some different fixed gap amplitudes $\Delta/\epsilon_F = 0.003, \ 0.01, \ 0.1$. 
$\phi \sim 0$ corresponds to the imaginary $s$-wave pairing fluctuation in the TSC and 
$\phi \sim \pi/2$ corresponds to the imaginary topological pairing fluctuation in the ordinary $s$-wave superconductor. 
In this model, for a realistic value of the gap amplitude $\Delta/\epsilon_F = o(10^{-1})$, 
fluctuations of the imaginary $s$-wave pairing order in the TSC induce dynamical axion, while
dynamical axion due to  fluctuations of the imaginary topological pairing order in the trivial superconductor is strongly suppressed. 
Because of this distinctive behavior of axion angle, we only consider the fluctuation of the imaginary $s$-wave pairing order in the TSC in the following sections. 
The ratio between the fluctuation $\Delta_s^{\mathrm{Im}}$ and the dynamical axion $\delta \theta$ is  
\begin{equation}\begin{split}
\delta \theta = 2 \pi c \frac{\Delta_s^{\mathrm{Im}}}{\Delta_p}
\end{split}\label{d-theta}\end{equation}
with $c = O(1)$. 

The consideration in this section is based on the mean-field Hamiltonian (\ref{HBdG}).
The $s$($p$)-wave order parameter field in (\ref{HBdG}) should be regarded as a fluctuating field in the case of the TSC (trivial superconductor), 
In the next section, we present more accurate and reliable calculations for order parameter fluctuations. 

\section{Dynamical axion in TSCs with $s$-wave pairing interaction}
\label{Dy-Axion}
As mentioned in the previous section, for spin-$1/2$ superconducting systems, 
the imaginary $s$-wave superconducting fluctuation in the TSC gives rise to (gravitational) dynamical axion field in condensed matter systems. 
In this section, we explore properties of dynamical axion fluctuations. 
The action of the dynamical axion is written by 
\begin{equation}\begin{split}
&S_{\mathrm{axion}}[\delta \theta] \\
&= J g^2 \int_0^{\beta} d \tau \int d^3 x \left[ (\partial_{\tau} \delta \theta)^2 + v_F^2 (\bm{\nabla} \delta \theta)^2 + m_{\theta}^2 (\delta \theta)^2 \right]
\end{split}\end{equation}
up to the second order of the fluctuation field $\delta \theta$, which amounts to the random-phase approximation (RPA). 
A parameter $g$ relates the dynamical axion field $\delta\theta$ and the superconducting fluctuation field $\delta\psi$, i.e. $\delta\theta=g\delta\psi$.
Here, $\delta\psi$ may be the fluctuation of the amplitude, or the phase of the superconducting order, or the mixture of them.
$g$ can be derived by expanding Eq.(\ref{axion}) in terms of $\delta\psi$ up to the 1st order.
The expressions of $J$, and the axion mass $m_{\theta}$ for a particular microscopic model are given in Appendix.
The axion mass $m_{\theta}$ controls fluctuation of the dynamical axion. 
For the  realization of dynamical axion, it is required that $m_{\theta}$ is smaller than the bulk superconducting gap $\Delta$. 
As a concrete example, we consider a $p$-wave TSC system similar to the B-phase of $^3$He.
We also assume that there is an $s$-wave-channel pairing interaction $U_s$ which yields s-wave superconducting fluctuations. \cite{Goswami-Roy}
In the case with inversion symmetry, pairing states with different parity can not be mixed with each other. 
Thus, in the $p$-wave pairing state, the $s$-wave superconducting order can not develop even when $U_s$ is nonzero.
However, fluctuation of s-wave pairing order is still possible.
On the other hand, in the case with broken inversion symmetry (i.e. noncentrosymmetric crystal structure), 
parity-mixing of $p$-wave pairing and $s$-wave pairing is induced by the antisymmetric SOI. \cite{bauer2012non}
In this case, the imaginary $s$-wave pairing fluctuation is not independent of the $p$-wave pairing fluctuation. 
Actually, dynamical axion emerges as a fluctuation of a phase difference between $s$-wave and $p$-wave superconducting orders {\it a la} the Leggett mode. \cite{leggett1966number, sharapov2002effective}
Generally, in the presence of two mean field superconducting order $\Delta_1$, $\Delta_2$, 
there are two superconducting fluctuations $\Delta_1 e^{i \theta_1}$, $\Delta_2 e^{i\theta_2}$. 
Intensity of the phase fluctuations is determined by a Coulomb interaction, which couples with the total phase fluctuation $\theta_1+\theta_2$, and 
a scattering channel between two Cooper pairs, which couples with the relative phase fluctuation (Leggett mode) $\theta_1-\theta_2$. 
It is noted that the Leggett mode has an advantage for the realization of dynamical axion because the relative phase avoids to get a plasma gap from a long-range Coulomb interaction, and thus dynamical axion can survive in the low-energy region. 
In the following, we consider two cases with and without the SOI, and compare properties of dynamical action for these two cases.
We will find that the SOI induced by broken inversion symmetry dramatically enlarges the parameter region in which dynamical axion
appears as a low-energy excitation.

\subsection{Model Hamiltonian}
We consider a toy model which has contact attractive interactions in both the $s$-wave and $p$-wave channels, and also the antisymmetric SOI. 
The Hamiltonian is 
\begin{equation}\begin{split}
H &= H_0 + H_s + H_p,  \\
\end{split}\end{equation}
\begin{equation}\begin{split}
H_0 
&= \sum_{\bm{k}} \left[ \left( \frac{k^2}{2 m} - \mu \right) \delta_{\alpha \beta } + \lambda \bm{k} \cdot \bm{\sigma}_{\alpha \beta} \right] c^{\dag}_{\bm{k}\alpha} c_{\bm{k}\beta}, 
\end{split}\end{equation}
\begin{equation}\begin{split}
H_s 
&= -\frac{U_s}{4 V}\sum_{\bm{k}\bm{k}'\bm{q}} \left( i \sigma_2 \right)_{\alpha\beta} \left( i \sigma_2 \right)_{\alpha'\beta'} \\
&\ \ \ \ \ \  c^{\dag}_{\bm{k}+\frac{\bm{q}}{2} \alpha} c^{\dag}_{-\bm{k}+\frac{\bm{q}}{2}\beta} c_{-\bm{k}'+\frac{\bm{q}}{2}\beta'}c_{\bm{k}'+\frac{\bm{q}}{2}\alpha'} , 
\end{split}\end{equation}
\begin{equation}\begin{split}
H_{p} 
&= -\frac{U_{p}}{4 V}\sum_{\bm{k}\bm{k}'\bm{q}} \left( \hat k \cdot \bm{\sigma} i \sigma_2 \right)_{\alpha\beta} \left( - i \sigma_2 \bm{\sigma} \cdot \hat k' \right)_{\alpha'\beta'} \\
&\ \ \ \ \ \ \ \ \ \ \ \ c^{\dag}_{\bm{k}+\frac{\bm{q}}{2} \alpha} c^{\dag}_{-\bm{k}+\frac{\bm{q}}{2}\beta} c_{-\bm{k}'+\frac{\bm{q}}{2}\beta'}c_{\bm{k}'+\frac{\bm{q}}{2}\alpha'} , 
\end{split}\end{equation}
where $\lambda$ is the SOI strength, $U_s$ and $U_p<0$ are, respectively, the $s$-wave and $p$-wave attractive interactions, and $\hat k = \bm{k}/k_F$ with $k_F$ the Fermi wave number for $\lambda = 0$. 
We assume that the $d$-vector of the $p$-wave pairing is the same as that of the B-phase of $^3$He, which  is a typical example of a class DIII TSC.
For simplicity, we ignore scatterings between the $s$-wave and $p$-wave pairs,
though such processes are generally allowed in the case without inversion symmetry.
This treatment is justified because scattering processes mixing different parity are higher order in terms of a parameter $\epsilon_{SO}/\epsilon_{F}$
where $\epsilon_{SO}$ is  the energy scale of the SOI, and $\epsilon_{F}$ is the Fermi energy, satisfying $\epsilon_{SO}/\epsilon_{F} \ll 1$ for most of real materials. \cite{bauer2012non} 
The mean field order parameter is written in the following form,
\begin{equation}\begin{split}
\Delta(\bm{k}) = \Delta_s i \sigma_2 + \Delta_p \hat k \cdot \bm{\sigma} i \sigma_2. 
\end{split}\end{equation}
$|\Delta_p| > |\Delta_s|$ corresponds to the TSC ($N = \pm 1$) and $|\Delta_p| < |\Delta_s|$ corresponds to the trivial superconductor ($N = 0$). 
The Hubbard-Stratonovich transformation leads to the partition function in the form,
\begin{equation}\begin{split}
Z = \int D \Delta D \Delta^* e^{-S[\Delta,\Delta^*]} , 
\end{split}\end{equation}
\begin{equation}\begin{split}
&S[\Delta,\Delta^*] \\
&= \int_0^{\beta} d\tau \int d^3 x \sum_{l=s,p} \frac{|\Delta_l(x)|^2}{U_l} - \frac{1}{2} \mathrm{Tr} \mathrm{ln} G^{-1}[\Delta, \Delta^*] , 
\end{split}\label{SD}\end{equation}
where
\begin{widetext}
\begin{equation}\begin{split}
&G^{-1}_{\alpha\beta}(\bm{k}_1\tau_1,\bm{k}_2\tau_2;\Delta,\Delta^*) \\
&= 
\begin{pmatrix}
- \partial_{\tau_1} \delta_{\bm{k}_1\bm{k}_2}\delta_{\alpha\beta} - H_{\alpha\beta}(\bm{k}_1,\bm{k}_2) & -\sum_{l=s,p} \sum_{\bm{q}} \Delta_{l}(\tau_1,\bm{q}) \varphi_{l,\bm{q}}([(\bm{k}_1\alpha)(-\bm{k}_2\beta)]) \\
-\sum_{l=s,p} \sum_{\bm{q}} \Delta^*_{l}(\tau_1,\bm{q}) \varphi^*_{l,\bm{q}}([(\bm{k}_2\beta)(-\bm{k}_1\alpha)]) & - \partial_{\tau_1} \delta_{\bm{k}_1\bm{k}_2}\delta_{\alpha\beta} + H_{\beta\alpha}(-\bm{k}_2,-\bm{k}_1) \\
\end{pmatrix} \delta(\tau_1-\tau_2) , 
\end{split}\end{equation}
\end{widetext}
with the normal part of the Hamiltonian,
\begin{equation}\begin{split}
H_{\alpha\beta}(\bm{k}_1,\bm{k}_2) = \left( \varepsilon_{\bm{k}_1} \delta_{\alpha\beta} + \lambda \hat k_1 \cdot \bm{\sigma}_{\alpha\beta} \right) \delta_{\bm{k}_1\bm{k}_2} , 
\end{split}\end{equation}
and the $s$-wave and $p$-wave pairing channel bases 
\begin{equation}\begin{split}
\varphi_{s,\bm{q}}([(\bm{k}_1\alpha)(\bm{k}_2\beta)]) = \delta_{\bm{q},\bm{k}_1+\bm{k}_2} (i \sigma_2)_{\alpha\beta} , 
\end{split}\end{equation}
\begin{equation}\begin{split}
\varphi_{p,\bm{q}}([(\bm{k}_1\alpha)(\bm{k}_2\beta)]) = \delta_{\bm{q},\bm{k}_1+\bm{k}_2} \left( \frac{\bm{k}_1-\bm{k}_2}{2 k_F} \cdot \bm{\sigma} i \sigma_2 \right)_{\alpha\beta} , 
\end{split}\end{equation}
where $\bm{q}$ is a momentum of center-of-mass motion of a Cooper pair. 
Here, we choose the Nambu spinor as $\Psi = (\psi_{\uparrow },\psi_{\downarrow },\psi^*_{\uparrow }, \psi^*_{\downarrow })^T$, and define
the Fourier transformation of the superconducting fluctuations $\Delta_l(\tau,\bm{x})$ by $\Delta_l(\tau,\bm{q}) = \frac{1}{V} \int d^3 x \Delta_l(\tau,\bm{x}) e^{-i \bm{q} \cdot \bm{x}}$.

\subsection{Case without spin-orbit interaction}
First, we consider the case without the SOI, i.e., $\lambda=0$, which corresponds to the case with inversion symmetry. 
The mean field free energy is obtained by setting $\Delta_l(\tau,\bm{x}) = \Delta_l = const.$:
\begin{widetext}
\begin{equation}\begin{split}
F[\Delta,\Delta^*] /V 
&= \frac{|\Delta_s|^2}{U_s} + \frac{|\Delta_{p}|^2}{U_{p}} 
- \frac{1}{2} \frac{1}{\beta V} \sum_{\omega_n\bm{k}} \ln \left[ \left( \omega_n^2 + \varepsilon^2_{\bm{k}} + |\Delta_s|^2 + (k/k_F)^2|\Delta_p|^2 \right)^2 - (k/k_F)^2 \left( \Delta_s^* \Delta_p + \Delta_s \Delta^*_p \right)^2 \right] . 
\end{split}\label{FEwithoutSOI}\end{equation}
\end{widetext}From this expression, it is found that the cases of the $\pi/2$ relative phase between $\Delta_s$ and $\Delta_p$, i.e., $\Delta_s^*\Delta_p+\Delta_s\Delta^*_p=0$, have the lowest free energy, 
which means the imaginary $s$-wave fluctuation is larger than the real $s$-wave fluctuation in the bulk $p$-wave superconducting phase. 
Within approximation of $k \sim k_F$ in the sum of $\bm{k}$ in the third term in (\ref{FEwithoutSOI}), we search for the free-energy minimum.
We find that for $U_s > U_p$, the $s$-wave order trivial phase $\Delta_s \neq 0, \Delta_p = 0$ is stabilized, 
while for $U_s < U_p$, the topological phase with the $p$-wave order, i.e. $\Delta_s = 0, \Delta_p \neq 0$, realizes. 
As mentioned in the previous section, here, 
we only consider fluctuations of $s$-wave pairing in the bulk $p$-wave superconductor in the case of $U_p>U_s$. 
The gap equation for $\Delta_p(T)$ is derived from the free energy minimum $\frac{\partial F[\Delta_p, \Delta_s=0]}{\partial \Delta_p^*} = 0$, 
\begin{equation}\begin{split}
\frac{1}{U_p} 
&= \frac{1}{\beta V} \sum_{\omega_n, \bm{k}} \frac{(k/k_F)^2}{\omega_n^2 + \varepsilon_{\bm{k}} + (k/k_F)^2 \Delta_p^2(T)}. 
\end{split}\label{Gap-s}\end{equation}

The action of the imaginary $s$-wave fluctuation is obtained by expanding (\ref{SD}) in terms of the imaginary part of the $s$-wave gap $\Delta_s^{\rm Im}$ within the Gaussian approximation:
\begin{equation}\begin{split}
&S[\Delta_s^{\mathrm{Im}}] \\
&= J \int_0^{\beta} d \tau \int d^3 x \left[ (\partial_{\tau} \Delta_s^{\mathrm{Im}})^2 + v_F^2 (\bm{\nabla} \Delta_s^{\mathrm{Im}})^2 + m_{\theta}^2 (\Delta_s^{\mathrm{Im}})^2 \right]
\end{split}\label{acion-s}\end{equation}
where $v_F = \partial_k \varepsilon_{\bm{k}} |_{k=k_F}$ is the Fermi velocity and the definition of $J$ and mass gap $m_{\theta}$ is given in Appendix. 
At zero temperature $T=0$, $m_{\theta}$ is 
\begin{equation}\begin{split}
m_{\theta}(T=0)
&= |\Delta_p| \sqrt{\frac{2}{\rho_0} \left( \frac{1}{U_s} - \frac{1}{U_p} \right)}
\end{split}\end{equation}
with $\rho_0 = 2mk_F/(2 \pi)^2$ the density of state per spin. 
The axion mass gap is determined by the distance from the topological phase transition point $U_s = U_p$. 
\begin{figure}[!]
 \begin{center}
  \includegraphics[width=\linewidth, trim=0cm 0cm 0cm 0cm]{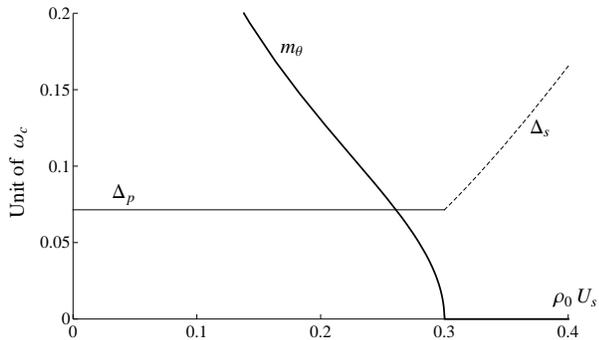}
 \end{center}
 \caption{The bulk superconducting gap $\Delta_p/\Delta_s$ and the axion mass gap $m_{\theta}$ in a unit of cutoff energy $\omega_c$ at zero temperature for a fixed value of $U_p$ plotted against $\rho_0 U_s$. The region $0 \leq \rho_0 U_s <0.3$ is the $p$-wave topological superconducting state, while
 the region  $0.3 \leq \rho_0 U_s $ is the trivial $s$-wave pairing state.
The first order topological phase transition occurs at $\rho_0 U_s = 0.3$ point. 
}
 \label{FIG2}
\end{figure}
In Fig.\ref{FIG2}, we show the bulk superconducting gap $\Delta_p$ and the axion mass gap $m_{\theta}$ at zero temperature for a fixed $\rho_0 U_p = 0.3$. 
It is noted that fluctuations of dynamical axion survive only in the very narrow parameter region in vicinity of the topological transition point $U_s = U_p$.
Thus, the dynamical axion scenario is unlikely to realize for centrosymmetric systems with no SOI , i.e. $\lambda=0$. 

\subsection{Case with spin-orbit interaction}
Next, let us consider the case with the finite SOI $\lambda \neq 0$. 
We fix $\lambda > 0$. 
Due to the parity mixing, the $s$-wave and the $p$-wave superconducting orders coexist. 
The mean field free energy is derived by setting $\Delta_l(\tau,\bm{x}) = \Delta_l = const.$, 
\begin{widetext}
\begin{equation}\begin{split}
& F[\Delta,\Delta^*] /V 
= \frac{|\Delta_s|^2}{U_s} + \frac{|\Delta_{p}|^2}{U_{p}} \\
& - \frac{1}{2} \frac{1}{\beta V} \sum_{\omega_n\bm{k}} \ln \left( \omega_n^2 + ( \varepsilon_{\bm{k}} + \lambda k)^2 + |\Delta_s + (k/k_F) \Delta_p|^2 \right)
- \frac{1}{2} \frac{1}{\beta V} \sum_{\omega_n\bm{k}} \ln \left( \omega_n^2 + ( \varepsilon_{\bm{k}} - \lambda k)^2 + |\Delta_s - (k/k_F) \Delta_p|^2 \right) . 
\end{split}\end{equation}
\end{widetext}
The second and third terms correspond to the contribution from the inner Fermi surface and the outer Fermi surface, respectively. 
The gap equation is derived from a free energy minimum condition $\frac{\partial F}{\partial \Delta_s^*} = \frac{\partial F}{\partial \Delta_p^*} = 0$, which leads to
\begin{widetext}
\begin{equation}\begin{split}
\frac{\Delta_s}{U_s} 
&= \frac{1}{2 \beta V} \sum_{\omega_n, \bm{k}} \frac{\Delta_s + (k/k_F) \Delta_p}{\omega_n^2 + ( \varepsilon_{\bm{k}} + \lambda k)^2 + |\Delta_s + (k/k_F) \Delta_p|^2 }
+ \frac{1}{2 \beta V} \sum_{\omega_n, \bm{k}} \frac{\Delta_s - (k/k_F) \Delta_p}{\omega_n^2 + ( \varepsilon_{\bm{k}} - \lambda k)^2 + |\Delta_s - (k/k_F) \Delta_p|^2 } , \\
\frac{\Delta_s}{U_p} 
&= \frac{1}{2 \beta V} \sum_{\omega_n, \bm{k}} \frac{\left\{ \Delta_s + (k/k_F) \Delta_p\right\} (k/k_F)}{\omega_n^2 + ( \varepsilon_{\bm{k}} + \lambda k)^2 + |\Delta_s + (k/k_F) \Delta_p|^2 }
- \frac{1}{2 \beta V} \sum_{\omega_n, \bm{k}} \frac{\left\{ \Delta_s - (k/k_F) \Delta_p\right\} (k/k_F)}{\omega_n^2 + ( \varepsilon_{\bm{k}} - \lambda k)^2 + |\Delta_s - (k/k_F) \Delta_p|^2 } . \\
\end{split}\label{Gap-p}\end{equation}
\end{widetext}
We denote the Fermi wave numbers for the inner/outer Fermi surface as $k_{F\pm}$ where $\pm$ represent the $\bm{k}$-dependent helicities for the spin degrees of freedom, i.e. $\bm{k} \cdot \sigma = \pm k$. 
When the normal energy dispersion is $\varepsilon_{\bm{k}} = \frac{k^2}{2 m} - \epsilon_F$, then $k_{F\pm}/k_F = \sqrt{1+\left(\frac{\lambda}{v_F}\right)} \mp \frac{\lambda}{v_F}$ with 
$k_F = \sqrt{2m \varepsilon_F}$ and $v_F=\frac{k_F}{m}$.
The bulk superconducting gaps for each Fermi surface are given by $\Delta_s + k_{F+}/k_F \Delta_p$ for the inner Fermi surface and $\Delta_s - k_{F-}/k_F \Delta_p$ for the outer Fermi surface. 
\begin{figure}[!]
 \begin{center}
  \includegraphics[width=\linewidth, trim=0cm 0cm 0cm 0cm]{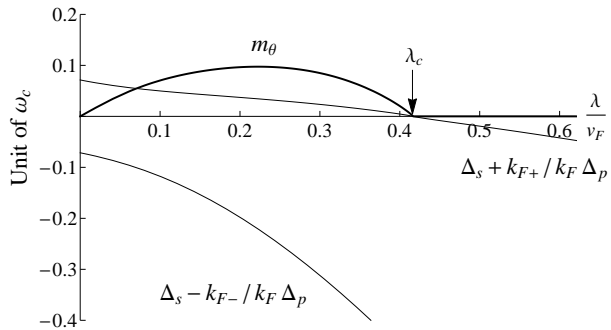}
 \end{center}
 \caption{The bulk superconducting gaps $\Delta_s \pm k_{F\pm}/k_F \Delta_p$ for inner/outer band and the axion mass gap $m_{\theta}$ 
in a unit of cut off energy $\omega_c$ at zero temperature for a fixed values of $\rho_0 U_s = 0.2$ and $\rho_0 U_p = 0.3$. 
$\lambda = \lambda_c$ is the topological phase transition point. 
}
 \label{FIG3}
\end{figure}
In Fig.\ref{FIG3}, we show the superconducting gap as a function of the SOI strength $\lambda$ for a fixed value of $U_s < U_p$ at zero temperature. 
The small $\lambda$ region corresponds to a topological phase since this region adiabatically connects to the pure $p$-wave topological phase. 
There is a topological phase transition point $\lambda_c$ at which the superconducting gap for the inner Fermi surface $\Delta_s + k_{F+}/k_F \Delta_p$ closes. 
At the phase transition point, the magnitude of the $s$-wave pairing interaction $U_s$ and that of the effective $p$-wave interaction $(k_{F+}/k_F) U_p$ are the same. 
The region for $\lambda > \lambda_c$ is the topologically trivial state since this region adiabatically connects to the ordinary $s$-wave superconductor $\Delta_p = 0$. 
To investigate properties of the dynamical axion induced by superconducting fluctuations, we only consider the topological region $\lambda<\lambda_c$. 

In the case with the SOI, because of the mixing of $\Delta_s$ and $\Delta_p$, the imaginary $s$-wave fluctuation is not independent of the $p$-wave superconducting fluctuation. 
As a matter of fact, the fluctuation inducing the dynamical axion is a relative phase fluctuation between $\Delta_s$ and $\Delta_p$. 
We define the relative phase $\theta_r$ by 
\begin{equation}\begin{split}
\Delta_s(\tau,\bm{x}) &= \Delta^{\mathrm{MF}}_s e^{\frac{i}{2} \theta_r(\tau,\bm{x})} , \\
\Delta_p(\tau,\bm{x}) &= \Delta^{\mathrm{MF}}_p e^{-\frac{i}{2} \theta_r(\tau,\bm{x})} . \\
\end{split}\end{equation}
The action of the relative phase fluctuation is given by (\ref{SD}) up to the Gaussian fluctuation around the mean field solution and within 
the long wave-length approximation, 
\begin{equation}\begin{split}
&S[\theta_r] \\
&= J \int_0^{\beta} d \tau \int d^3 x \left[ (\partial_{\tau} \theta_r)^2 + v_F^2 (\bm{\nabla} \theta_r)^2 + m_{\theta}^2 (\theta_r)^2 \right]
\end{split}\end{equation}
where $v_F = \partial_k \varepsilon_{\bm{k}} |_{k=k_F}$ is Fermi velocity and the definition of $J$ and mass gap $m_{\theta}$ is given in Appendix. 
In Fig.\ref{FIG3}, we show the axion mass gap $m_{\theta}$ as a function of $\lambda$ at zero temperature. 
For sufficiently small and realistic values of $\lambda/v_F$, $m_{\theta}$ is much smaller than the superconducting gap, and hence, the dynamical axion realizes as a low-energy excitation. 
Here, we have ignored parity non-conserving pairing interactions such as the scattering between $s$-wave Cooper pairs and $p$-wave Cooper pairs. 
Such channels induce an additional gap in the relative phase fluctuation. 
However, they are higher order in terms of the SOI,\cite{bauer2012non} and thus, the effect on the axion mass gap is small. 

Here we remark some important features which are different from the case without the SOI. 
As shown in Fig.\ref{FIG2}, in the absence of the SOI, the axion mass gap $m_{\theta}$ increases quite rapidly as the distance from the topological phase transition point $U_s = U_p$ increases. 
Thus, in this case, the realization of the dynamical axion requires a fine tuning of the interaction strength $U_s$ and $U_p$.  
On the other hand, in the case with the SOI, due to the finite gaps of both $s$-wave and $p$-wave superconducting orders, 
the relative phase can fluctuate, and for sufficiently small $\lambda$, the axion mass gap is smaller than the bulk superconducting gaps. 
The region where the dynamical axion survives always exists for any interaction strength satisfying $U_s < U_p$. 
Moreover the relative phase does not couple with a long-range Coulomb interaction. Therefore, it is free from acquiring a plasma gap. 
Hence, noncentrosymmetric superconductors with the antisymmetric SOI 
are significantly promising systems for the realization of dynamical axion in TSCs.

\section{Dynamical axion phenomena in topological superconductors}
\label{Dy-Phe}
In this section, we discuss the physically observable phenomena raised by the dynamical axion via the gravitoelectromagnetic $\theta$ term (\ref{FE}). 
In the condensed matter context, in the case with the U(1) particle number conservation, 
the dynamical axion $\theta$ couples with electromagnetic fields via the topological $\theta$ term $S_{\theta} = \frac{\alpha}{32 \pi^2} \int dt d^3 x \theta \epsilon^{\mu\nu\rho\sigma} F_{\mu\nu} F_{\rho\sigma}$, 
which induces novel phenomena such as axionic polariton under an applied uniform magnetic field \cite{li2010dynamical} and magnetic instability under an applied uniform electrostatic field \cite{ooguri2012instability}. 
An important question is, then, what dynamical axion phenomena are in topological superconductors or superfluids for which U(1) symmetry is broken. 

We discuss the dynamical axion phenomena on the basis of the topological action (\ref{FE}), 
in which mechanical rotation couples with dynamical axion. 
Let us consider an increase of the moment of inertia raised by the dynamical axion fluctuation under finite temperature gradient. 
The real time action per unit volume for the rotation $\Omega = \dot \phi$ and dynamical axion $\delta \theta$ is given by 
\begin{equation}\begin{split}
S[\phi, \delta \theta] 
&= \int d t \frac{1}{2} I \dot \phi^2 
+ J g^2 \int d t \left( \dot{\delta \theta}^2 - m_{\theta}^2 \delta \theta^2 \right) \\
& - \frac{k_{\mathrm{B}}^2 T \nabla T}{12 \hbar v^2} \int d t \dot \phi \delta \theta  , 
\end{split}\end{equation}
where $I$ is the inertia moment per unit volume, and we assume that the temperature gradient $\bm{\nabla} T$ is parallel to $\bm{\Omega}$.  
Integrating out the dynamical axion fluctuation, we get an effective action for rotation only, 
\begin{equation}\begin{split}
S_{\mathrm{eff}}[\phi] 
&= \int \frac{d \omega}{2 \pi} \frac{I_{\mathrm{eff}}(\omega)}{2} \omega^2 |\phi(\omega)|^2 
\end{split}\end{equation}
with
\begin{equation}\begin{split}
I_{\mathrm{eff}}(\omega) = I + \frac{1}{2 J g^2 m_{\theta}^2} \left( \frac{k_{\mathrm{B}}^2 T \nabla T}{12 \hbar v^2} \right)^2 \frac{1}{1-(\omega/m_{\theta})^2}.
\end{split}\end{equation}
The result implies that
the mechanical rotation excites the dynamical axions via the topological coupling (\ref{FE}), 
which increase the inertia moment for a low-frequency region $\omega < m_{\theta}$. 
Let us estimate the order of the increase of the inertia moment. 
We approximate $J$, $g$, and $v$ by 
\begin{equation}\begin{split}
J \sim \rho = \frac{k_F^3}{8 \pi^3 \epsilon_F}, \ 
g \sim \pi , \ 
v \sim \frac{\Delta}{\hbar k_F} , 
\end{split}\label{Ap}\end{equation}
and the bare inertia moment per unit volume by 
\begin{equation}\begin{split}
I \sim M/L_z
\end{split}\end{equation}
with $M$ the system total mass and $L_z$ the system size in the direction parallel to the temperature gradient. 
Then we obtain, 
\begin{equation}\begin{split}
&\frac{I_{\mathrm{eff}}(\omega)-I}{I} \\
&\sim \frac{\pi}{36} 
\left( \frac{k_B T}{\Delta} \right)^2
\left( \frac{k_B L_z \nabla T}{\Delta} \right)^2
\left( \frac{\hbar}{m_{\theta}} \right)^2
\frac{\epsilon_F k_F}{M L_z}
\frac{1}{1-(\omega/m_{\theta})^2}. 
\end{split}
\label{moment}
\end{equation}
The prefactor of Eq.(\ref{moment}) is extremely small.  However, in the case of ac mechanical rotation, i.e. a shaking motion with a finite frequency $\omega$,  as $\omega$ approaches the resonance frequency $m_{\theta}$, this effect becomes observable.

\section{conclusion}
\label{Conc}

In this paper, we discussed dynamical axion in topological superconductors and superfluids in terms of the gravitoelectromagnetic-type topological action (\ref{FE}), 
in which the axion and mechanical rotation are coupled under finite temperature gradient. 
Here, we stress that the microscopic derivation of the topological action (\ref{FE}) for non-quantized axion angle $\theta$ and its fluctuation $\delta \theta$ is still an open issue. 
We have assumed the topological action (\ref{FE}) and $\theta$ is determined by the Chern-Simons 3 form, $\theta = 2 \pi \int CS_3(\mathcal{A})$. 
Under this assumption, the superconducting fluctuations which shift axion angle defined by the Chern-Simons 3 form give rise to the dynamical axion. 
The superconducting fluctuations inducing the dynamical axion in the topological superconductors are the time-reversal broken $s$-wave fluctuations in the absence of the antisymmetric SOI, 
and the relative phase fluctuation (Leggett mode) between $p$-wave and $s$-wave orders in the presence of the antisymmetric SOI. 
We found that the SOI breaking inversion symmetry enlarges the parameter region in which the dynamical axion fluctuation appears as a low-energy excitation, 
since the magnitude of the relative phase fluctuation is determined by the coupling strength between $s$-wave and $p$-wave Cooper pairs, i.e., the strength of the SOI. 
We proposed that the dynamical axion fluctuation increases the moment of inertia. 
If the rotation frequency $\omega$ is close to the dynamical axion fluctuation mass $m_{\theta}$, this effect is observable.


\begin{acknowledgments}
K.S. thanks T. Kimura for helpful discussion. 
This work was supported by the Grant-in-Aids for Scientific
Research from MEXT of Japan [Grants No. 23540406, No. 25220711, and No. 25103714 (KAKENHI on Innovative Areas "Topological Quantum Phenomena")]. 
K.S. is supported by a JSPS Fellowship for Young Scientists. 

\end{acknowledgments}



\begin{widetext}
\appendix
\section{A derivation of the RPA action of superconducting fluctuations}
In this appendix, we give the detailed derivation of the action for the superconducting fluctuations that give rise to dynamical axion.

\subsection{Case without SOI}
We consider a bulk $p$-wave superconductor $\Delta_p \neq 0$. 
We can expand in terms of fluctuations of $\Delta_s$ around the mean field order $\Delta_p$. From (\ref{SD}), up to the second order in $\Delta_s^{\mathrm{Im}}$ we get
\begin{equation}\begin{split}
S[\Delta^{\mathrm{Im}}_s] 
&= V \int_0^{\beta} d\tau \sum_{\bm{q}} \frac{1}{U_s} \Delta^{\mathrm{Im}}_s(\tau,-\bm{q}) \Delta^{\mathrm{Im}}_s(\tau,\bm{q})  
+ \frac{1}{2} \mathrm{Tr} \left[ G_{0} \hat \Delta \right] + \frac{1}{4} \mathrm{Tr} \left[ G_{0} \hat \Delta \right]^2 + O([\Delta_s^{\mathrm{Im}}]^3)
\end{split}\label{action-s'}\end{equation}
with
\begin{equation}\begin{split}
G_{0}^{-1}(\bm{k}_1\tau_1,\bm{k}_2\tau_2)
&= 
\begin{pmatrix}
- \partial_{\tau_1} - \varepsilon_{\bm{k}_1} & - \Delta_p \hat k_1 \cdot (\bm{\sigma} i \sigma_2) \\
- \Delta_p \hat k_1 \cdot (-i \sigma_2 \bm{\sigma}) & - \partial_{\tau_1} + \varepsilon_{-\bm{k}_1} \\
\end{pmatrix} \delta_{\bm{k}_1\bm{k}_2} \delta(\tau_1-\tau_2) , \\
\end{split}\end{equation}
\begin{equation}\begin{split}
\hat \Delta(\bm{k}_1\tau_1,\bm{k}_2\tau_2)
&= 
\begin{pmatrix}
0 & \sum_{\bm{q}} i \Delta^{\mathrm{Im}}_s(\tau_1,\bm{q}) \delta_{\bm{q},\bm{k}_1-\bm{k}_2} (i \sigma_2) \\
\sum_{\bm{q}} -i \Delta^{\mathrm{Im}}_s(\tau_1,-\bm{q}) \delta_{\bm{q},\bm{k}_2-\bm{k}_1} (-i \sigma_2) & 0 \\
\end{pmatrix} \delta(\tau_1-\tau_2) .
\end{split}\end{equation}
Here, note that $[ \Delta^{\mathrm{Im}}_s(\tau,\bm{q}) ]^* = \Delta^{\mathrm{Im}}_s(\tau,-\bm{q})$. 
The second term in (\ref{action-s'}) vanishes.  Then, we have, 
\begin{equation}\begin{split}
S[\Delta^{\mathrm{Im}}_s] 
&= \beta V \sum_{\Omega_m.\bm{q}} \Delta^{\mathrm{Im}}_s(-\Omega_m,-\bm{q}) \left[ \frac{1}{U_s} + \Pi^{\mathrm{Im}\mathrm{Im}}_{ss} (\Omega_m,\bm{q})\right] \Delta^{\mathrm{Im}}_s(\Omega_m,\bm{q}) 
\end{split}\label{action-s}\end{equation}
with
\begin{equation}\begin{split}
\Pi^{\mathrm{Im}\mathrm{Im}}_{ss} (\Omega_m,\bm{q})
&= \frac{1}{4 \beta V} \sum_{\omega_n\bm{k}} \mathrm{tr} \left[ G_0(\omega_n+\Omega_m/2,\bm{k}+\bm{q}/2) (-\tau_1 \sigma_2) G_0(\omega_n-\Omega_m/2,\bm{k}-\bm{q}/2) (-\tau_1 \sigma_2) \right]
\end{split}\label{Pi-s}\end{equation}
where $\Omega_m = 2 \pi m/\beta$ is  the boson Matsubara frequency, and $\bm{\tau} = (\tau_1,\tau_2,\tau_3)$ represent Pauli matrices for Nambu space, and $G_0(\omega_n,\bm{k})$ is the mean field Green's function
\begin{equation}\begin{split}
\left[ G_0(\omega_n,\bm{k}) \right]^{-1} = \begin{pmatrix}
i\omega_n - \varepsilon_{\bm{k}} & - \Delta_p \hat k \cdot (\bm{\sigma} i \sigma_2) \\
- \Delta_p \hat k \cdot (-i \sigma_2 \bm{\sigma}) & i\omega_n + \varepsilon_{-\bm{k}} \\
\end{pmatrix}. 
\end{split}\end{equation}From low-frequency and long wave-length approximations,
we get the action Eq. (\ref{acion-s}) with 
\begin{equation}\begin{split}
J = \frac{\partial^2 \Pi^{\mathrm{Im}\mathrm{Im}}}{\partial \Omega_m^2} (0,\bm{0}) 
&= \frac{1}{\beta V} \sum_{\omega_n\bm{k}} \frac{1}{\left\{ \omega_n^2 + \varepsilon_{\bm{k}} + (k/k_F)^2 \Delta_p^2 \right\}^2} , 
\end{split}\label{J-s}\end{equation}
and 
\begin{equation}\begin{split}
J m_{\theta}^2 
&= \frac{1}{U_s} + \Pi^{\mathrm{Im}\mathrm{Im}}_{ss} (0,\bm{0}) \\
&= \frac{1}{U_s} - \frac{1}{\beta V} \sum_{\omega_n\bm{k}} \frac{1}{\omega_n^2 + \varepsilon_{\bm{k}} + (k/k_F)^2 \Delta_p^2} . \\
\end{split}\label{m-s}\end{equation}
We approximate $k/k_F \sim 1$ and $\frac{1}{V}\sum_{\bm{k}} f(\varepsilon_{\bm{k}}) \cong \rho_0 \int_{-\omega_c}^{\omega_c}d \xi f(\xi) $ with a cutoff $\omega_c$ in (\ref{Gap-s}), (\ref{J-s}) and (\ref{m-s}),
and consider the zero temperature limit.
Then,  we obtain,
\begin{equation}\begin{split}
\Delta_p(T=0)
&= \omega_c \sinh \left( \frac{1}{\rho_0 U_p} \right) , \ 
J = \frac{\rho_0}{2 \Delta_p^2} , \ 
J m_{\theta}^2 = \frac{1}{U_s} - \frac{1}{U_p} . 
\end{split}\end{equation}

\subsection{Case with SOI}
Next, we consider the case with finite SOI, which induces parity-mixing of $\Delta_s$ and $\Delta_p$. 
The action of superconducting fluctuations up to the second order is 
\begin{equation}\begin{split}
S[\delta \Delta] 
&= \int_0^{\beta} d\tau \int d^3 x \left( \frac{( \delta |\Delta_s| )^2}{U_s} + \frac{( \delta |\Delta_p| )^2}{U_p} \right)
+ \frac{1}{2} \mathrm{Tr} \left[ G_{0} \delta \hat \Delta \right] + \frac{1}{4} \mathrm{Tr} \left[ G_{0} \delta \hat \Delta \right]^2 + O(\delta \Delta^3),
\end{split}\label{action-p'}\end{equation}
with 
\begin{equation}\begin{split}
G_{0}^{-1}(\bm{k}_1\tau_1,\bm{k}_2\tau_2)
&= 
\begin{pmatrix}
- \partial_{\tau_1} - \varepsilon_{\bm{k}_1} & - \Delta_s i \sigma_2 - \Delta_p \hat k_1 \cdot (\bm{\sigma} i \sigma_2) \\
\Delta_s i \sigma_2 - \Delta_p \hat k_1 \cdot (-i \sigma_2 \bm{\sigma}) & - \partial_{\tau_1} + \varepsilon_{-\bm{k}_1} \\
\end{pmatrix} \delta_{\bm{k}_1\bm{k}_2} \delta(\tau_1-\tau_2) \\
\end{split}\label{G0-p}\end{equation}
the mean field Green's function. 
The first term in (\ref{action-p'}) comes from the amplitude fluctuation of superconducting order. 
Since the fluctuation inducing the dynamical axion is the relative phase $\theta_r$ between $\Delta_s$ and $\Delta_p$, we deal with only relative phase fluctuations. 
For the finite SOI strength $\lambda$, the relative phase mode is massive, so we can expand around the fixed mean field relative phase, 
\begin{equation}\begin{split}
\delta \Delta_s(\tau,\bm{x}) 
&= \Delta_s e^{\frac{i}{2} \theta_r(\tau,\bm{x})} - \Delta_s
\cong \Delta_s \left( \frac{i}{2} \theta_r(\tau,\bm{x}) - \frac{1}{8} \theta_r^2(\tau,\bm{x}) \right) , 
\end{split}\end{equation}
\begin{equation}\begin{split}
\delta \Delta_p(\tau,\bm{x}) 
&= \Delta_p e^{-\frac{i}{2} \theta_r(\tau,\bm{x})} -\Delta_p
\cong \Delta_p \left( - \frac{i}{2} \theta_r(\tau,\bm{x}) - \frac{1}{8} \theta_r^2(\tau,\bm{x}) \right) .
\end{split}\end{equation}
Here $\Delta_s$ and $\Delta_p$ include the mean field relative phase $\theta_r^{\mathrm{MF}} = 0$ or $\pi$. 
Using the Fourier transformation from $\bm{x}$ to $\bm{q}$, we get 
\begin{equation}\begin{split}
\delta \Delta_s(\tau,\bm{q}) 
\cong \Delta_s \left( \frac{i}{2} \theta_r(\tau,\bm{q}) - \frac{1}{8} \sum_{\bm{q}'} \theta_r(\tau,\bm{q}) \theta_r(\tau,\bm{q}-\bm{q}') \right) , 
\end{split}\end{equation}
\begin{equation}\begin{split}
\delta \Delta_p(\tau,\bm{q}) 
&= \Delta_p e^{-\frac{i}{2} \theta_r(\tau,\bm{x})} -\Delta_p
\cong \Delta_p \left( - \frac{i}{2} \theta_r(\tau,\bm{q}) - \frac{1}{8} \sum_{\bm{q}'} \theta_r(\tau,\bm{q}) \theta_r(\tau,\bm{q}-\bm{q}') \right) .
\end{split}\end{equation}
Then $\delta \hat \Delta$ which includes the relative phase fluctuations is written as 
\begin{equation}\begin{split}
& \delta \hat \Delta(\bm{k}_1\tau_1,\bm{k}_2\tau_2) \\
&= 
\begin{pmatrix}
0 & \sum_{\bm{q}} \delta \Delta_s(\tau_1,\bm{q}) \delta_{\bm{q},\bm{k}_1-\bm{k}_2} (i \sigma_2) \\
\sum_{\bm{q}} \delta \Delta^*_s(\tau_1,\bm{q}) \delta_{\bm{q},\bm{k}_2-\bm{k}_1} (-i \sigma_2) & 0 \\
\end{pmatrix} \delta(\tau_1-\tau_2) \\
& + 
\begin{pmatrix}
0 & \sum_{\bm{q}} \delta \Delta_p(\tau_1,\bm{q}) \delta_{\bm{q},\bm{k}_1-\bm{k}_2} \left( \frac{\bm{k}_1+\bm{k}_2}{2 k_F} \cdot \bm{\sigma} i \sigma_2 \right) \\
\sum_{\bm{q}} \delta \Delta^*_p(\tau_1,\bm{q}) \delta_{\bm{q},\bm{k}_2-\bm{k}_1} \left( - i \sigma_2 \bm{\sigma} \cdot \frac{\bm{k}_1+\bm{k}_2}{2 k_F} \right) & 0 \\
\end{pmatrix} \delta(\tau_1-\tau_2) \\
&= \sum_{\bm{q}} \theta_r(\tau,\bm{q}) 
\begin{pmatrix}
0 & \frac{i}{2} \left\{ \Delta_s - \Delta_p \left( \frac{\bm{k}_1+\bm{k}_2}{2 k_F} \cdot \bm{\sigma} \right) \right\} i \sigma_2  \\
-\frac{i}{2} (-i \sigma_2) \left\{ \Delta_s - \Delta_p \left( \frac{\bm{k}_1+\bm{k}_2}{2 k_F} \cdot \bm{\sigma} \right) \right\} & 0 \\
\end{pmatrix} \delta_{\bm{q},\bm{k}_1-\bm{k}_2} \delta(\tau_1-\tau_2) \\
&+ \sum_{\bm{q} \bm{q}'} \theta_r(\tau,\bm{q}) \theta_r(\tau,\bm{q}-\bm{q}')
\begin{pmatrix}
0 & -\frac{1}{8} \left\{ \Delta_s + \Delta_p \left( \frac{\bm{k}_1+\bm{k}_2}{2 k_F} \cdot \bm{\sigma} \right) \right\} i \sigma_2  \\
-\frac{1}{8} (-i \sigma_2) \left\{ \Delta_s + \Delta_p \left( \frac{\bm{k}_1+\bm{k}_2}{2 k_F} \cdot \bm{\sigma} \right) \right\} & 0 \\
\end{pmatrix} \delta_{\bm{q},\bm{k}_1-\bm{k}_2} \delta(\tau_1-\tau_2) . \\
\end{split}\label{dD-p}\end{equation}
Note that $\theta^*_r(\tau,\bm{q}) = \theta_r(\tau,-\bm{q})$. From (\ref{action-p'}), (\ref{G0-p}) and (\ref{dD-p}) we obtain the RPA action for the relative phase $\theta_r$,  
\begin{equation}\begin{split}
S[\theta_r]
&= \beta V \sum_{\Omega_m \bm{q}} \theta_r(-\Omega_m,-\bm{q}) \left[ C + \Pi_r(\Omega_m,\bm{q}) \right] \theta_r(\Omega_m,\bm{q}) , 
\end{split}\end{equation}
where
\begin{equation}\begin{split}
C = \frac{1}{\beta V} \sum_{\omega_n\bm{k}} \mathrm{tr} \left[ G_0(\omega_n,\bm{k}) 
\begin{pmatrix}
0 & -\frac{1}{8} \left\{ \Delta_s + \Delta_p \left( \frac{\bm{k}}{k_F} \cdot \bm{\sigma} \right) \right\} i \sigma_2  \\
-\frac{1}{8} (-i \sigma_2) \left\{ \Delta_s + \Delta_p \left( \frac{\bm{k}}{k_F} \cdot \bm{\sigma} \right) \right\} & 0 \\
\end{pmatrix} \right]
\end{split}\end{equation} 
and 
\begin{equation}\begin{split}
\Pi_r(\Omega_m,\bm{q})
&= \frac{1}{\beta V} \sum_{\omega_n\bm{k}} \mathrm{tr} \Bigg[ G_0(\omega_n+\Omega_m/2,\bm{k}+\bm{q}/2) 
\begin{pmatrix}
0 & \frac{i}{2} \left\{ \Delta_s - \Delta_p \left( \frac{\bm{k}}{k_F} \cdot \bm{\sigma} \right) \right\} i \sigma_2  \\
-\frac{i}{2} (-i \sigma_2) \left\{ \Delta_s - \Delta_p \left( \frac{\bm{k}}{k_F} \cdot \bm{\sigma} \right) \right\} & 0 \\
\end{pmatrix} \\
& \ \ \ \ \ \ \ \ \ \ \ \ \ \ \ \ \ \ \ 
G_0(\omega_n-\Omega_m/2,\bm{k}-\bm{q}/2) \begin{pmatrix}
0 & \frac{i}{2} \left\{ \Delta_s - \Delta_p \left( \frac{\bm{k}}{k_F} \cdot \bm{\sigma} \right) \right\} i \sigma_2  \\
-\frac{i}{2} (-i \sigma_2) \left\{ \Delta_s - \Delta_p \left( \frac{\bm{k}}{k_F} \cdot \bm{\sigma} \right) \right\} & 0 \\
\end{pmatrix} \Bigg] \\
\end{split}\end{equation}
with 
\begin{equation}\begin{split}
G^{-1}_0(\omega_n,\bm{k}) = 
\begin{pmatrix}
i\omega_n - \varepsilon_{\bm{k}} & - \Delta_s i \sigma_2 - \Delta_p \hat k \cdot (\bm{\sigma} i \sigma_2) \\
\Delta_s i \sigma_2 - \Delta_p \hat k \cdot (-i \sigma_2 \bm{\sigma}) & i\omega_n + \varepsilon_{-\bm{k}} \\
\end{pmatrix} . 
\end{split}\end{equation}
For low-frequency and long wave-length regions, we obtain, 
\begin{equation}\begin{split}
S[\theta_r] 
&= J \int_0^{\beta} d \tau \int d^3 x \left[ (\partial_{\tau} \theta_r)^2 + v_F^2 (\bm{\nabla} \theta_r)^2 + m_{\theta}^2 \theta_r^2 \right]
\end{split}\label{action-p''}\end{equation}
with 
\begin{equation}\begin{split}
J = \frac{\partial^2 \Pi_r}{\partial \Omega_m^2} (0,\bm{0}),
\end{split}\label{m-p1}
\end{equation}
\begin{equation}\begin{split}
J m_{\theta}^2 = C + \Pi_r(0,\bm{0}). 
\end{split}\label{m-p}\end{equation}
Using the approximation $k/k_F \sim 1$ and $\frac{1}{V}\sum_{\bm{k}} f(\varepsilon_{\bm{k}}) \cong \rho_{\pm} \int_{-\omega_c}^{\omega_c} d \xi f(\xi) $ with a cutoff $\omega_c$ and 
$\rho_{\pm}$ the density of state per spin for the inner/outer Fermi surface, and considering the zero temperature limit, we simplify the gap equation (\ref{Gap-p}) into the form, 
\begin{equation}\begin{split}
\frac{1}{\rho U_s} \frac{\Delta_s}{\omega_c} + \frac{1}{\rho U_p} \frac{\Delta_p}{\omega_c} 
&= \frac{\rho_+}{\rho_0} \frac{\Delta_s + (k_{F+}/k_F) \Delta_p}{\omega_c} \sinh \left( \frac{\omega_c}{\left| \Delta_s + (k_{F+}/k_F) \Delta_p \right|} \right) ,  \\
\frac{1}{\rho U_s} \frac{\Delta_s}{\omega_c} - \frac{1}{\rho U_p} \frac{\Delta_p}{\omega_c} 
&= \frac{\rho_-}{\rho_0} \frac{\Delta_s - (k_{F-}/k_F) \Delta_p}{\omega_c} \sinh \left( \frac{\omega_c}{\left| \Delta_s - (k_{F-}/k_F) \Delta_p \right|} \right) . \\
\end{split}\end{equation}
Also, Eqs. (\ref{m-p1}) and (\ref{m-p}) are 
\begin{equation}\begin{split}
J = \frac{1}{16} \left[ \rho_+ \left( \frac{\Delta_s - (k_{F+}/k_F) \Delta_p}{\Delta_s + (k_{F+}/k_F) \Delta_p} \right)^2 + \rho_- \left( \frac{\Delta_s + (k_{F-}/k_F) \Delta_p}{\Delta_s - (k_{F-}/k_F) \Delta_p} \right)^2 \right], 
\end{split}\end{equation}
\begin{equation}\begin{split}
J m^2_{\theta}
&= \frac{1}{2} \frac{\Delta_s (k_{F+}/k_F) \Delta_p}{\Delta_s + (k_{F+}/k_F) \Delta_p} \left( \frac{\Delta_s}{U_s} + \frac{\Delta_p}{U_p} \right) 
 - \frac{1}{2} \frac{\Delta_s (k_{F-}/k_F) \Delta_p}{\Delta_s - (k_{F-}/k_F) \Delta_p} \left( \frac{\Delta_s}{U_s} - \frac{\Delta_p}{U_p} \right). 
\end{split}\end{equation}

\end{widetext}



\end{document}